\documentclass[11pt]{article}
\usepackage{epsfig,graphicx, amssymb}
\evensidemargin  \oddsidemargin
\usepackage{amsmath}

\begin{document}

\begin{center}
{\Large
 Vibration of the Duffing Oscillator: Effect of Fractional Damping
\\ ~\\ ~\\  
Marek Borowiec and Grzegorz Litak
\\~
Department of Applied Mechanics, Technical University of
Lublin, Nadbystrzycka 36, PL-20-618 Lublin, Poland
\\~\\~
Arkadiusz Syta
\\~
Department of Applied Mathematics, Technical University
of
Lublin,
Nadbystrzycka 36, PL-20-618 Lublin, Poland,}
\end{center}

\vspace{2cm}

\begin{abstract}
 We have applied the Melnikov criterion to examine a global homoclinic bifurcation and transition 
to chaos 
 in a case of the
Duffing system with nonlinear fractional damping and external excitation. 
 Using perturbation methods we have found a critical forcing amplitude above which the system may 
 behave chaotically.
 The results have been verified by numerical simulations using standard nonlinear tools as 
 Poincare maps and a Lyapunov 
 exponent. Above the critical Melnikov amplitude $\mu_c$, which is the sufficient 
condition of a global 
homoclinic bifurcation, 
we have observed the region with  a transient chaotic motion. \\ \\


\end{abstract}

\newpage

\section{Introduction}
\begin{figure}[htb]
\centerline{
\epsfig{file=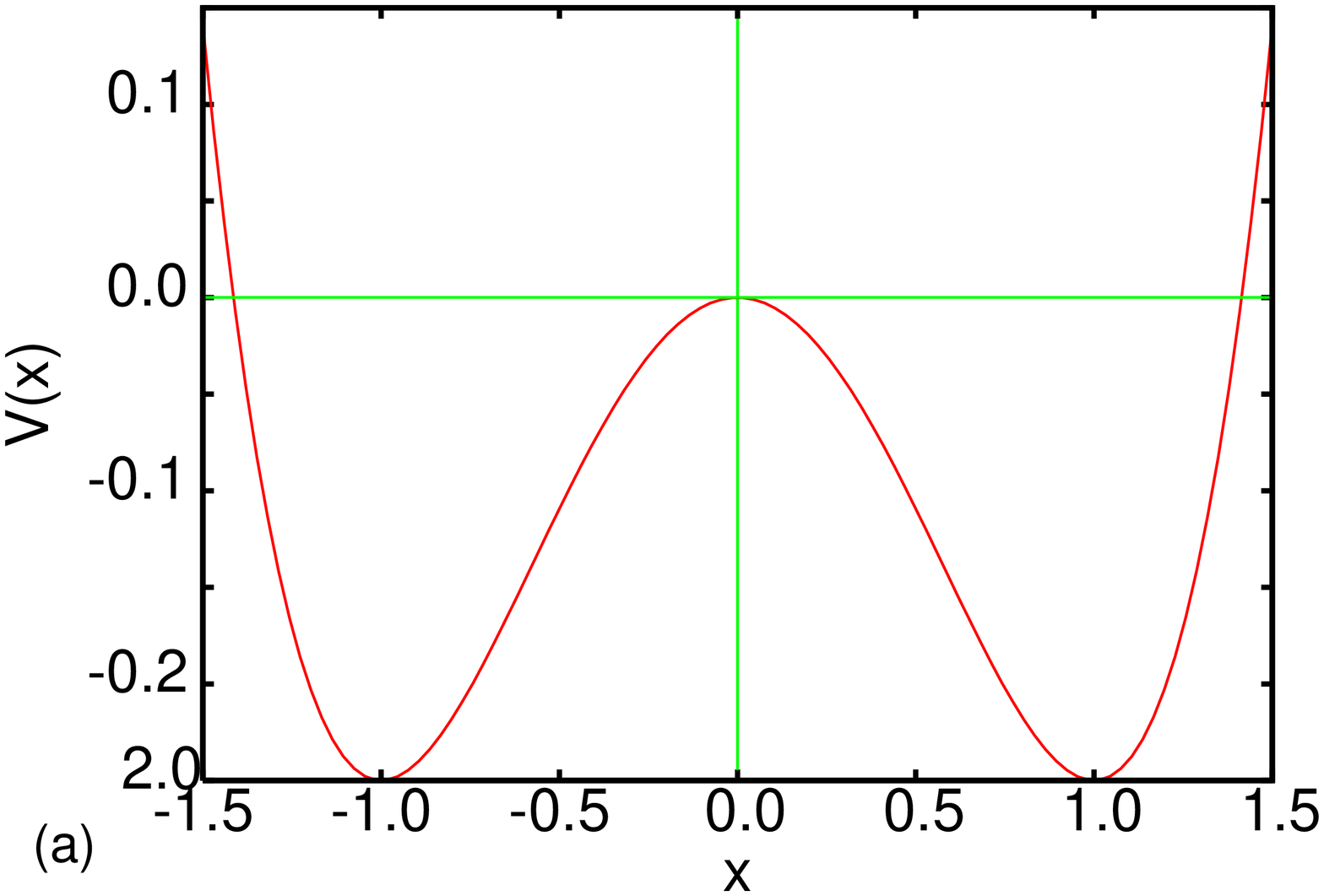,width=9.5cm,angle=-90}}

\centerline{
\epsfig{file=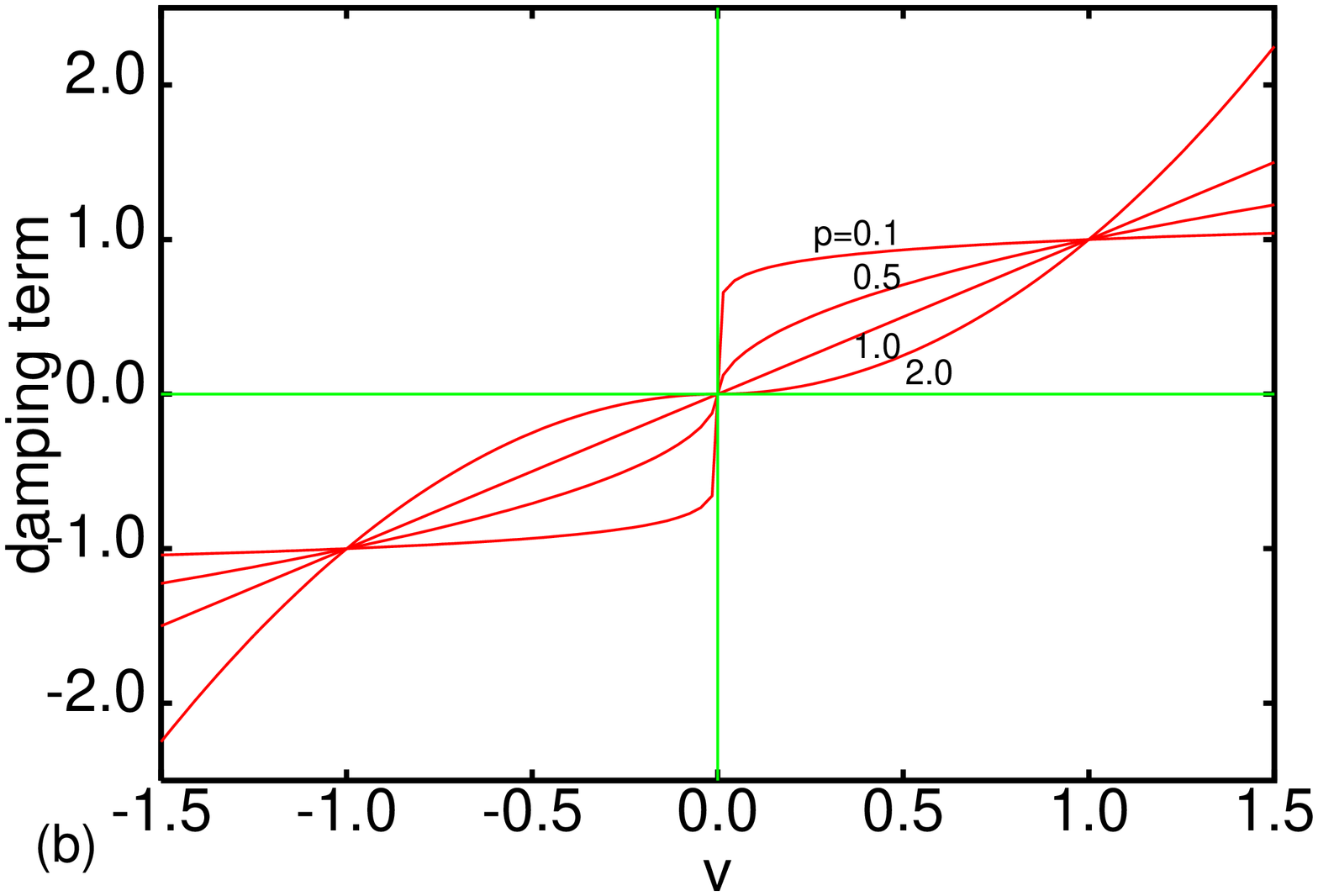,width=9.5cm,angle=-90}
}
\caption{ \label{fig1}
External potential $V(x)$ for $\delta=-1$ and $\gamma=1$  (Fig. \ref{fig1}a); and
velocity ($v=\dot x$) dependence of a damping term
${\rm dpt}(\dot{x})/\alpha$
for different $p$
normalized by
$\alpha$ (Fig. \ref{fig1}b).}
 \end{figure}

A nonlinear oscillator with the Duffing term but linear damping is   one of the simplest 
systems leading to chaotic 
motion studied originally by Ueda
\cite{Ueda1979,Ueda1980,Ueda1981}. 
The problem of  its nonlinear vibrations  
has attracted researchers from various 
fields of research; from natural science  \cite{Moon1979,Zalalutdinov2003} and mathematics 
\cite{Guckenheimer1983} to mechanical \cite{Szemplinska1993,Szemplinska1995,Moon1987} and 
electrical 
engineering \cite{Ueda1979,Ueda1980,Ueda1981}.
This system, for a negative linear part of stiffness, shows homoclinic orbits, 
and the transition to chaotic vibration can be treated 
analytically
by the Melnikov method. Note that such a treatment has been already
 performed successfully to selected problems with various potentials  
\cite{Guckenheimer1983,Wiggins1990}.
In spite of fact that a large bibliography is devoted to vibration of a single realization 
[1-12]
or set of 
coupled Duffing oscillators  \cite{Warminski1999,Warminski2000,Lifshitz2003,Maccari2002,Maccari2003}
with numerous 
modifications to potential and
forcing parts,
the problem of nonlinear
 damping in chaotically vibrating system has been focused shortly \cite{Trueba2000}. 
The fractionally damped Duffing's equation, for a single well potential with positive 
linear part, has been studied in detail by Padovan and Sawicki 
\cite{Padovan1998}. They were especially focused on the determination of the influence 
of fractional 
damping 
on the 
frequency amplitude response.
Some additional 
insight into this problem can be also found   
in the context of self excitation effects
 \cite{Litak1999,Warminski1999,Maccari2002,Maccari2003,Li2004}.
On the other hand, in the paper of Trueba {\em et al.} \cite{Trueba2000}, 
the systematic discussion
 on square and cubic damping effects on global homoclinic bifurcations in the 
Duffing system was given given while
Awrejcewicz and Holicke applied the Melnikov theory to Duffing system in 
a dry friction limit \cite{Awrejcewicz1999}.
In the present paper we reexamine the problem of nonlinear 
damping including  fractional 
damping cases by both; the analytical Melnikov method as well as numerical simulations.
Namely we are going to study a single degree of freedom nonlinear oscillator with the Duffing 
potential and fractional damping:
%eq1
\begin{equation}
\label{eq1}
\ddot{x} +\alpha \dot{x} \left| \dot{x} \right|^{p-1}
+ \delta x +\gamma x^3=\mu \cos{ \omega t},
\end{equation}
where $x$ is displacement and $\dot{x}$ velocity, respectively, while
the external potential force $F_x$ and corresponding potential $V(x)$ (Fig. 
\ref{fig1}a) are defined; 
%eq2
\begin{equation}
\label{eq2}
 -\delta x -\gamma x^3 =F_x=-\frac{{\rm d} V}{{\rm d} x},~~~~V(x)=\frac{\delta x^2}{2} + 
\frac{\gamma x^4}{4}.
\end{equation}

Here the non-linear damping term is defined by the exponent $p$:
%eq3
\begin{equation} 
\label{eq3}
{\rm dpt}(\dot{x})=\alpha \dot{x} \left| \dot{x} \right|^{p-1}.
\end{equation}
We have plotted the above function versus velocity ($v=\dot{x}$) in Fig. 
\ref{fig1}b for few values of $p$.
Note that,
the case
$p \rightarrow 0$ (see $p=0.1$ in Fig. \ref{fig1}b  for a relatively small
velocity) mimics the
dry
friction phenomenon \cite{Brockley1970,Ibrahim1994}.

\section{Melnikov Analysis and Numerical Simulations}
We start our analysis from the
unperturbed Hamiltonian
$H^0$: 
%eq4
\begin{equation}
\label{eq4}
H^0= \frac{v^2}{2} + V(x).
\end{equation}
The potential  function $V(x)$ (Fig. \ref{fig1}a) has the local peak at
the saddle point $x=0$.
Existence of this point with a horizontal tangent makes possible
homoclinic bifurcations to take place. This includes transitions from regular to
chaotic solutions. To study effects of damping and excitation, we apply small perturbations around the homoclinic orbits.
This implies using a small parameter $\epsilon$ in 
the Eq. \ref{eq1} with perturbation terms. 
Equation \ref{eq1} can be simultaneously uncoupled into two differential equations of the first 
order:
%eq5
\begin{eqnarray}
\label{eq5}
\dot{x} &=& v \\
\dot{v} &=& -\epsilon \tilde{\alpha} v \left| v \right|^{p-1}
- \delta x -\gamma x^3+\epsilon \tilde{\mu} \cos{ \omega t}, \nonumber
\end{eqnarray}
where $\epsilon \tilde{\alpha}=\alpha$ and $\epsilon \tilde{\mu}=\mu$, respectively.

\begin{figure}[htb]
\centerline{
\epsfig{file=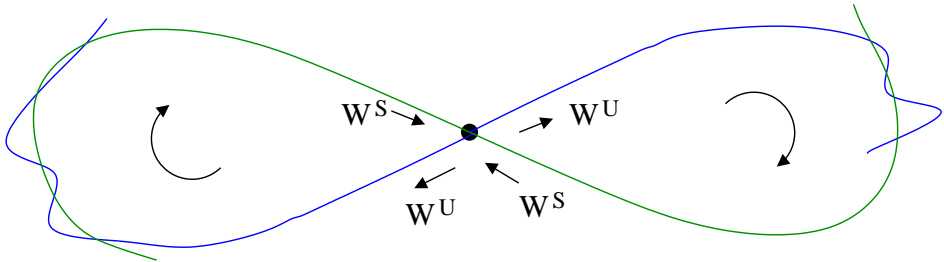,width=12.5cm,angle=0}
}
\caption{ \label{fig2} Perturbed homoclinic orbits (Eq. \ref{eq8}):
stable and unstable manifolds
($W^S$ and $W^U$).}
\vspace{10cm}
~
 \end{figure}

At the saddle point $x=0$, for an unperturbed system (Fig. \ref{fig1}a), the system velocity reaches 
zero in velocity $v=0$ 
(for infinite time $t=\pm \infty$) so the
total energy has only its
potential part.
Transforming Eqs. 2,4 for a chosen nodal energy ($E=0$) 
and for $\delta < 0$, $\gamma > 0$ we get the following expression for velocity:
%eq6
\begin{equation}
\label{eq6}
v= \frac{{\rm d} x}{{\rm d} t} =
\sqrt{2 \left(-
\frac{\delta x^2}{2} - \frac{\gamma
x^4}{4}\right)}.
\end{equation}
Now one can perform integration  over $x$:
%eq7
\begin{equation}
\label{eq7}
t-t_0= \pm \int  \frac{ {\rm d} x}{x\sqrt{
-\delta  - \frac{\gamma
x^2}{2}}},
\end{equation}
where $t_0$ represents an integration constant.
Finally, we get so called homoclinic orbits (Fig. \ref{fig2}):
%eq8
\begin{eqnarray}
x^* &=& \pm \sqrt{\frac{-2 \delta}{\gamma}}~~ 
\frac{1}{\cosh \left( \sqrt{-\delta} (t-t_0) \right)}
\nonumber \\
v^* &=& \pm 
\sqrt{\frac{2}{\gamma}}~\delta~~\frac{\tanh \left( \sqrt{-\delta} (t-t_0) \right)}{\cosh \left( 
\sqrt{-\delta} (t-t_0) \right)},
\label{eq8}
\end{eqnarray}
where '$+$' and '$-$' signs are related to left-- and right--side orbits, respectively (Fig. 
\ref{fig2}).
Note, the central saddle point $x_0=0$ is reached in time $t$
corresponding to $+\infty$ and $-\infty$, respectively.

In case of perturbed orbits $W^S$ (a stable manifold) and $W^U$
(an unstable manifold) the
distance between them is given
by
the Melnikov function ${\rm M}(t_0)$:
%eq9
\begin{equation}
\label{eq9}
{\rm M}(t_0) = \int_{- \infty}^{ + \infty}  h( x^*, v^*)  \wedge g( x^*,
v^*) {\rm d} t
\end{equation}
where the corresponding differential form $h$ means the gradient of unperturbed
Hamiltonian (Eq. \ref{eq4}):
%eq10
\begin{equation}
\label{eq10}
h = \left(\delta x^* + \gamma (x^*)^3\right) {\rm d} x  + v^* {\rm d}v,
\end{equation}
while $g$ is a perturbation form  (Eq. \ref{eq5}) to the same Hamiltonian:
%eq11
\begin{equation}
\label{eq11}
g = \left( \tilde{ \mu} \cos{\omega t} - \tilde{\alpha} v^* \left| v^* \right|^{p-1} \right) 
{\rm d}x.
\end{equation}
All differential forms are
defined on homoclinic orbits $(x,v)=(x^*,v^*)$ (Eq. \ref{eq8}).

Condition for a  global homoclinic transition, corresponding to a
horse-shoe
type of stable and unstable manifolds
cross-section (Fig. \ref{fig2}), can be written as:
%eq12
\begin{equation}
\label{eq12}
{\displaystyle \bigvee_{t_0}}
~~~ {\rm M}(t_0)=0 {\rm ~~ and ~~}
\frac{\partial {\rm M}(t_0)}{\partial t_0} \neq 0.
\end{equation}
Evaluating the corresponding integral (Eq. \ref{eq9})
after some lengthly algebra the last condition (Eq. \ref{eq12}) yields to
a critical value of excitation amplitude \cite{Moon1979, Trueba2000}
$\mu_c$:

%eq13
\begin{equation}
\label{eq13}
\mu_c= \alpha \frac{2^{p/2}}{\pi \omega} 
\frac{(-\delta)^{p+1/2}}{\gamma^{p/2}} {\rm B} 
\left( 
\frac{p+2}{2},\frac{p+1}{2}\right) 
\cosh \left( \frac{\pi \omega}{2 \sqrt{-\delta}}\right),
\end{equation}
where the B$(r,s)$ is the Euler Beta function:

%eq14
\begin{equation}
\label{eq14}
{\rm B}(r,s)=\frac{\Gamma (r)\Gamma (s)}{\Gamma (r + s)}
\end{equation}

and $\Gamma(n)$ denotes the Euler Gamma function:

%eq15
\begin{equation}
\label{eq15}
\Gamma (n + 1)= n\Gamma (n).
\end{equation}

\begin{figure}[htb]
\centerline{
\epsfig{file=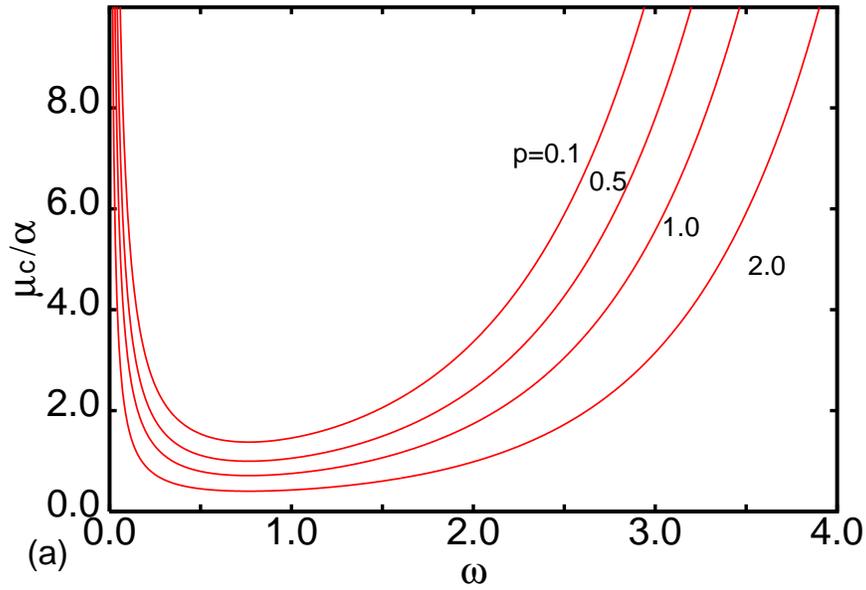,width=9.5cm,angle=-90}}

\centerline{
\epsfig{file=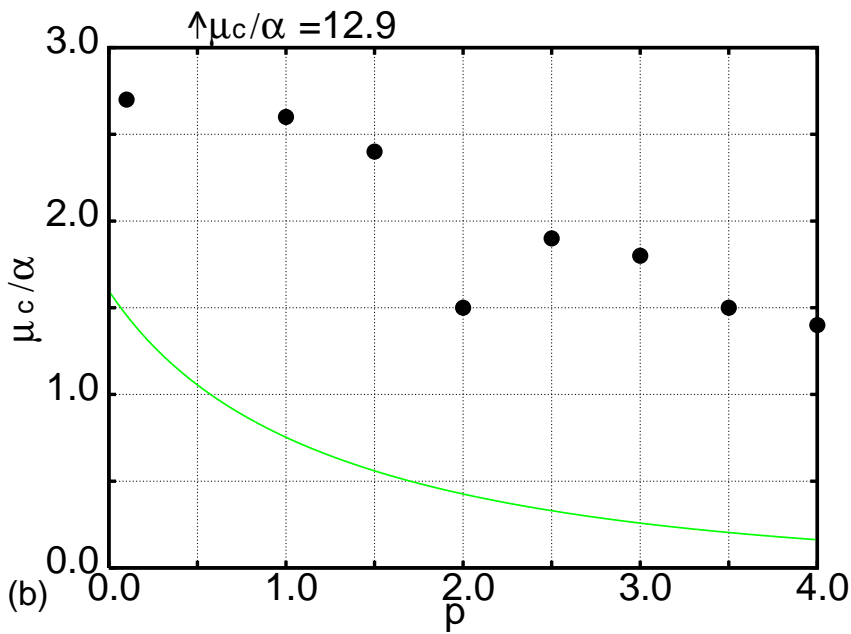,width=9.5cm,angle=-90}
}
\caption{ \label{fig3} Critical value of excitation $\mu_c$ versus $\omega$ (Fig.
\ref{fig3}a) and
exponent $p$ (Fig. \ref{fig3}b) for $\delta=-1$, $\gamma=1$. Points  in Fig. 
\ref{fig3}b shows 
numerical
results of transitions to chaos
obtained by Lyapunov exponents.}
 \end{figure}

\begin{figure}[htb]
\centerline{
\epsfig{file=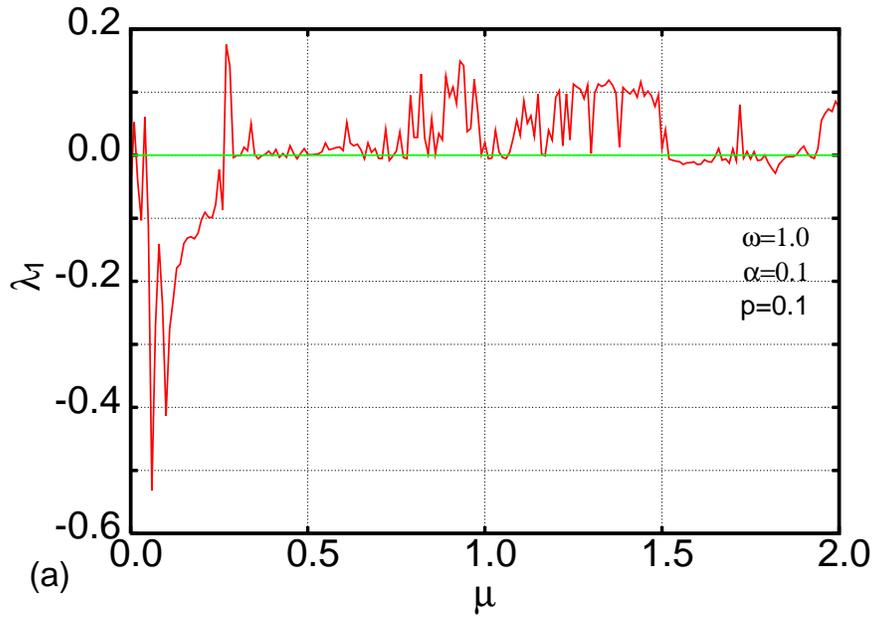,width=9.5cm,angle=-90}}             

\centerline{
\epsfig{file=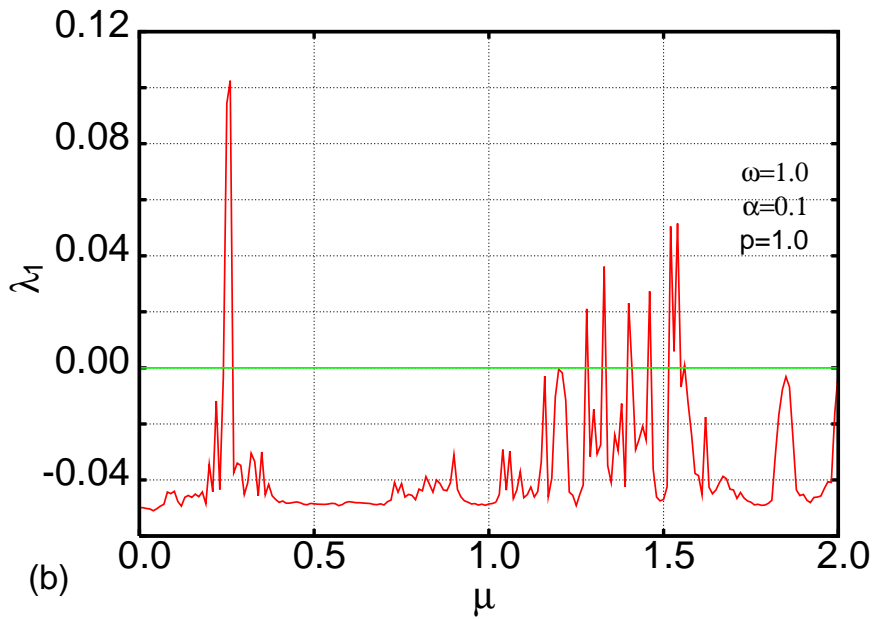,width=9.5cm,angle=-90}
}
\caption{ \label{fig4} Lyapunov exponents  for $\delta=-1$, $\gamma=1$ and $p=0.1$ 
(Fig. \ref{fig4}a), $p=1.0$ (Fig. \ref{fig4}b).}
 \end{figure}

In Fig. \ref{fig3}a we plotted the results of Melnikov analysis for a critical amplitude $\mu_c/\alpha$ 
for 
few values of $p$ while $\delta=-1$, $\gamma=1$. 
For $\mu > \mu_c$ the system can transit to chaotic vibrations.
In the next figure (Fig. \ref{fig3}b) we plotted $\mu_c$ versus $p$. Here  we  have also compared the 
results of analytical investigations represented by a solid line 
with
numerical simulations represented by circle points. 
The numerical results can give a local criterion of chaotic vibration appearance, namely a 
positive  value of the largest Lyapunov exponent $\lambda_1$ \cite{Wolf1985}.
Note that in all cases (Fig. \ref{fig3}b) points are above the line.  
This can be associate with the fact that the Lyapunov  exponent is a sufficient criterion 
of system transition to chaos while the Melnikov condition is only necessary one. 
Thus we can conclude that 
both sets of results: analytical and 
numerical are in good agreement.
The examples of the largest Lyapunov exponent $\lambda_1$ versus the excitation amplitude $\mu$ 
are plotted 
in Figs. \ref{fig4}a and b. 
In our numerical code we started calculations from the same initial conditions $(x_0,v_0)=(0.45,0.1)$ for every 
new value of $\mu$.  
One can clearly see the point of $\lambda_1$ sign change. This indicates transition to chaotic vibrations. Note that for $p=0.1$ we have got some small but positive values of
$\lambda_1$ for weak excitation ($\mu$ close to 0). When we examined this case with more details we observed that amplitude of vibrations was very small.
Appearance of small positive 
value of $\lambda_1$ seams to be a combined effect  of physical phenomenon of dry friction and our assumed 
numerical 
integration procedure of constant time steps $\Delta t$(in most cases we 
used $\Delta t = 2\pi/(500 \omega)$) . 
It is known from the studies of a stick-slip 
phenomenon \cite{Galvanetto1999,Leine2000}  that in such cases integration should be performed very 
carefully. 
Indeed, making time steps relatively smaller (50 times smaller) we checked that it is
sufficient to reduce the positive value around $\mu=0$ leaving, in principle, unchanged that 
$\lambda_1$ which appears at $\mu \approx 0.27$. To reduce this artificial effect of small positive 
values of $\lambda_1$ and assuming an efficient algorithm (not so much time consuming) we decided to put
a minimal threshold  of $\lambda_1=0.1$ above which the system was identified in a chaotic state (see 
Figs. \ref{fig3}b, \ref{fig4}a and \ref{fig4}b).

\begin{figure}[htb]
\centerline{
\epsfig{file=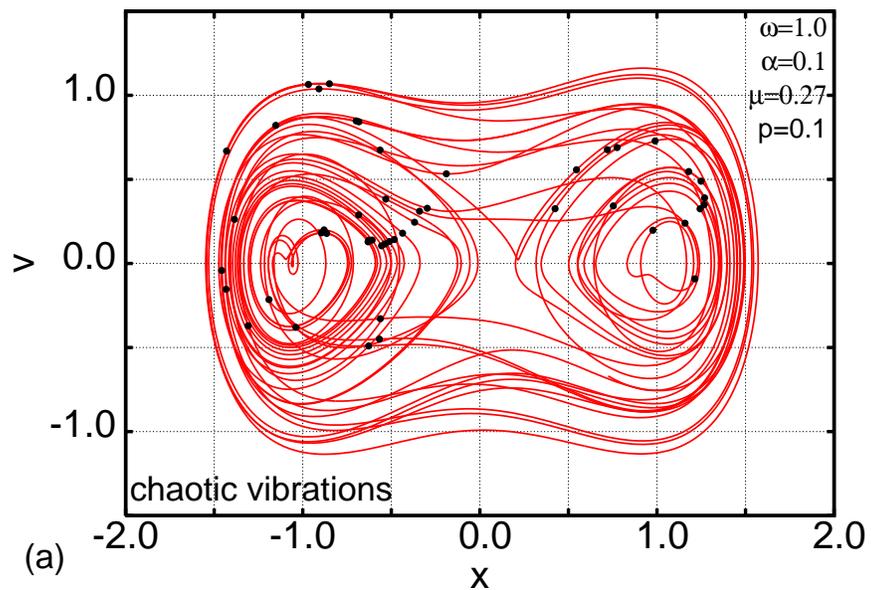,width=9.5cm,angle=-90}}             

\centerline{
\epsfig{file=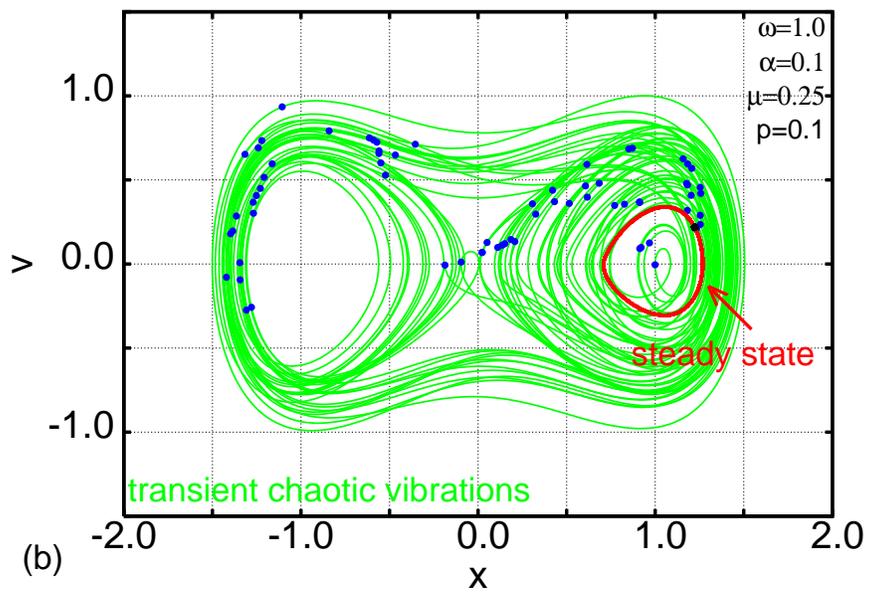,width=9.5cm,angle=-90}
}
\caption{\label{fig5} Phase portraits and Poincare maps   for $\delta=-1$, $\gamma=1$ and $\mu=0.27$ 
(Fig. 
\ref{fig5}a), 
$\mu=0.25$ (Fig. \ref{fig5}b).}
 \end{figure}

For a given value of $p$ ($p=0.1$) we show (in Fig. \ref{fig5}a for  $\mu=0.27$
 and Fig. \ref{fig5}b for $\mu=0.25$) typical phase portraits (represented by lines) and 
simultaneously 
Poincare maps 
 (represented by points). Note that in Fig. \ref{fig5}a we plotted trajectories for $t \in [1200,1500]$, 
in 
arbitrary units of $1/\omega$,  
 while for Fig. \ref{fig5}b we found transient behaviour. To show differences  we plotted corresponding 
trajectories for $t \in [0,1200]$  
 and $t \in [1200,1500]$ in various colours. The characteristic time  $t=1200$ has been chosen in such a 
way  to reach by our system
a steady state.
It is easy to see that Fig. \ref{fig5}a represents a non-periodic vibration state ($\mu=0.27$) 
just after the system transition to 
a chaotic 
motion 
 (see function $\lambda_1$ versus $\mu$ in Fig. \ref{fig4}a). The corresponding Lyapunow exponent 
$\lambda_1=0.175$ is positive indicating 
 on exponential divergence. In spite of a small number of Poincare stroboscopic points we can see a 
general view of a strange 
chaotic attractor.
On the other hand Fig. \ref{fig5}b shows system behaviour  state for a smaller excitation amplitude 
($\mu=0.25$). 
 In this case the initial trajectories follow the strange attractor complex geometry but after long 
enough time ($t=1200$) collapse to 
a steady state  motion. 
Here the corresponding 
Lyapunov exponent $\lambda_1=-0.023$ is negative which means 
a convergent motion. 

\begin{figure}[htb]
\centerline{
\epsfig{file=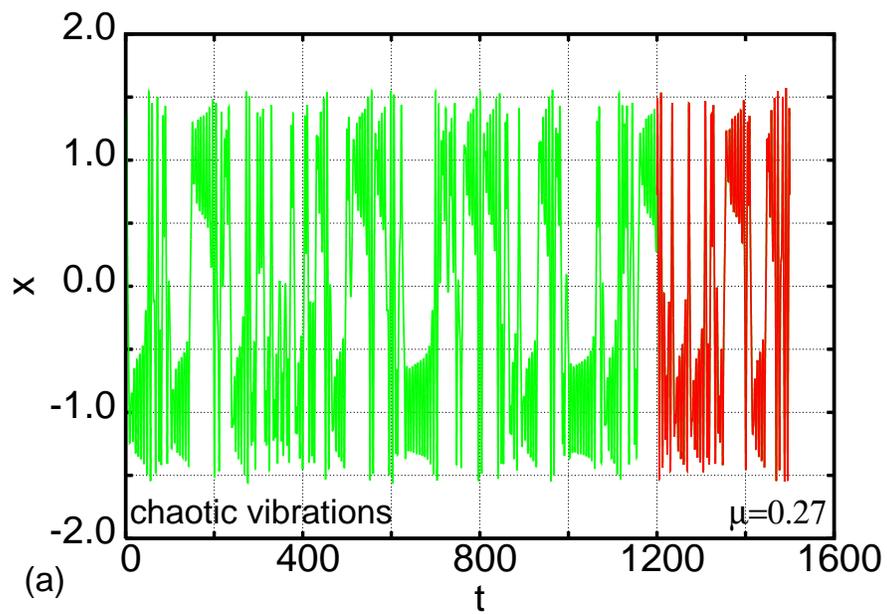,width=9.5cm,angle=-90}}             

\centerline{
\epsfig{file=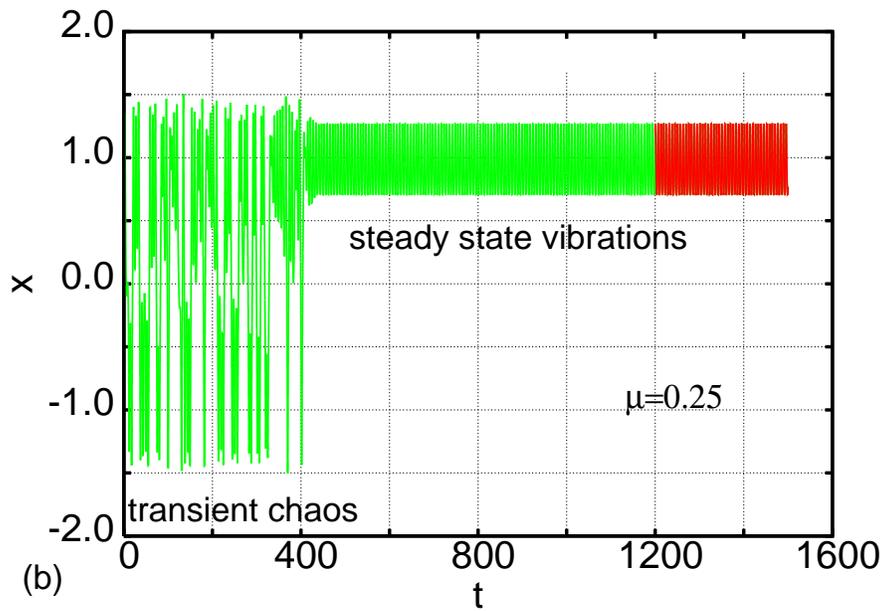,width=9.5cm,angle=-90}
}
\caption{ \label{fig6} Time histories   for $\delta=-1$, $\gamma=1$ and $\mu=0.27$ (Fig. \ref{fig6}a), 
$\mu=0.25$ 
(Fig. \ref{fig6}b).}
 \end{figure}

    For better clarity we also plotted  the corresponding time histories (Figs. \ref{fig6}a and 
b) using 
    the same colour as in Figs. \ref{fig5}a and b. One can easy note that for $\mu=0.27$ (Figs. 
\ref{fig5}a and 
\ref{fig6}a) we 
    have chaotic
     motion of the system while for $\mu=0.25$ (Figs. \ref{fig5}b and \ref{fig6}b) after some time of 
chaotic transient
      motion with
     a large amplitude of oscillations and jumping between potential wells we got steady state
     regular vibrations of much smaller amplitude located in a single potential well.

\section{Summary and Conclusions} 
We have examined criteria for transition to
chaotic
vibrations
in the Duffing system with a  damping term ${\rm dpt} (v)=v|v|^{p-1}$  described by a term 
fractional 
exponent $p$. We were interested especially in  $p < 1.0$ ($p=0.1$) representing an important 
case  which can be associated with a dry friction phenomenon \cite{Ibrahim1994}. The critical value of 
excitation amplitude $\mu$ above which the system vibrate chaotically has been estimated, in 
the first step, by 
means of the Melnikov method and later confirmed by calculating corresponding  Lyapunov exponent. 

The Melnikov method, is sensitive to a
global homoclinic bifurcation 
and gives a necessary condition for excitation amplitude $\mu=\mu_{c1}$ system in its transition 
to 
chaos \cite{Guckenheimer1983,Wiggins1990}. On the other hand the largest Lyapunov exponent \cite{Wolf1985}, measuring the local 
exponential 
divergences 
of particular phase portrait trajectories gives a sufficient  condition $\mu=\mu_{c2}$  for this 
transition
which has obviously a higher value of the excitation amplitude $\mu=\mu_{c2} > 
\mu_{c1}$.

Above the Melnikov transition predictions ($\mu > \mu_{c1}$) we have got transient chaotic vibrations 
\cite{Wiggins1990,Szemplinska1993,Szemplinska1995}
as we expected
drifting to a regular steady state away the fractal attraction regions separation  boundary. 
This is typical behaviour of the system which undergoes
global homoclinic bifurcation.       
In the region of resonance the amplitude of vibration depends on the damping exponent in a nontrivial 
way \cite{Padovan1998}. In particalar amplitude response rised for small fractional exponents 
($p<1$). Namely, such increase can be identified 
as an 
analogues of 
increasing system  response on stronger external forcing. Thus one may expect that a chaotic threshold 
would be reached 
easier in that case. In fact, comparing Figs. \ref{fig4}a and b, we observe larger Lyapunov exponents 
for a 
fractional
$p$ ($p=0.1$ in  Fig. \ref{fig4}a), however
from our Melnikov theory analysis we draw a different conclusion. The 
cases 
for $p<1$ (Figs. \ref{fig3}a-b) 
seem to be more stable against transition to chaos then the $p=1$ case. 
This interesting 
contradiction 
should be clearyfied by further simultaneous studies of 
local bifurcations and resonances in the above system. 

For small $p$ and in the limit of a small velocity we have found a region of system parameters leading 
to an interesting dry friction 
effect. Note, in our approach this limit has been obtained naturally on the same footing as other 
realizations. This enabled us to compare results of various damping terms (Figs. \ref{fig1}b and 
\ref{fig2}b). 
However to get reliable results in that region one must improve a simple numerical algorithm.  
We have 
not studied this effect 
systematically in this paper leaving it for a future  publication. 

The present approach can be used to generalize models of magnetorheological 
dampers in novel studies of their inflence on vehicle dynamics 
\cite{Li2004}.
To reduce harmful vibrations one can consider application of dampers composed of 
fractional terms.

\section*{Acknowledgments}
This paper has been partially  supported by the  Polish Ministry of 
Science and Informatization. Authors would like to thank 
organizers of the Conference {\em Recent Advances in Nonlinear 
Mechanics} in  Aberdeen (2005) for giving them an 
opportunity to preset these results. GL would like to thank the Max Planck 
Institute for the Physics of Complex Systems for hospitality.

%Figure1
\medskip

%Figure2
\medskip

%Figure3
\medskip

%Figure4
\medskip

%Figure5
\medskip

%Figure6   
\medskip

\end{document}